\documentclass[sigconf]{acmart}

\usepackage{draftwatermark} 

\begin{document}

\title{A replication of a controlled experiment with two STRIDE variants}


\author{Winnie Mbaka}
\email{w.mbaka@vu.nl}
\affiliation{%
  \institution{Vrije Universiteit Amsterdam}
  \country{The Netherlands}
}
\author{Katja Tuma}
\email{k.tuma@vu.nl}
\affiliation{%
  \institution{Vrije Universiteit Amsterdam}
  \country{The Netherlands}
}



\begin{abstract}
To avoid costly security patching after software deployment, security-by-design techniques (e.g., STRIDE threat analysis) are adopted in organizations to root out security issues before the system is ever implemented.
Despite the global gap in cybersecurity workforce and the high manual effort required for performing threat analysis, organizations are ramping up threat analysis activities.
However, past experimental results were inconclusive regarding some performance indicators of threat analysis techniques thus practitioners have little evidence for choosing the technique to adopt.
To address this issue, we replicated a controlled experiment with STRIDE. 
Our study was aimed at measuring and comparing the performance indicators (productivity and precision) of two STRIDE variants (element and interaction).
We conclude the paper by comparing our results to the original study.
\end{abstract}



\settopmatter{printfolios=true}
\maketitle
\SetWatermarkText{Pre-print} 
\SetWatermarkScale{0.5} 


\section{Introduction}
\label{sec:introduction}


Security-by-design techniques~\cite{DSS+2009,STM2017}, aim to forefront some security effort (to the design phase) and prevent costly security fixes to software in production.
For instance, threat analysis~\cite{tatam2021review,ling2020systematic} is used to scrutinize the software architecture for potential security issues.
STRIDE~\cite{shostack2014threat} is a popular threat analysis technique developed by Microsoft.
The outcomes of such analyses include adopting mitigations early-on or planning architectural refactoring to limit the use of external (possibly vulnerable) libraries and to minimize the attack surface of the software system.

Indeed, organizations are ramping up their investments in performing architectural threat and risk analysis (latest BSIMM study gathered data from 128 organizations and reports an increase by more than 65\%~\cite{BSIMM12:online}).
However, previous research has shown that architectural threat analysis requires a high manual effort~\cite{scandariato2015descriptive} and demands the involvement of security and domain experts~\cite{cruzes2018challenges}.
But this is hampered by a significant shortage of the security workforce.
This trend has been reported across sectors and globally.
According to CyberSeek, the United States faced a shortage of about 314,000 (44\% of the total, 716,000,  employed cybersecurity workforce at the time) security professionals in January 2019~\cite{cybersec2019}. 
Similar gap was observed for the EU security workforce~\cite{blavzivc2021cybersecurity}.

Empirical evidence of threat analysis performance indicators is a crucial piece of the puzzle.
First, such evidence supports security experts in understanding the trade-offs between the myriad of existing threat analysis techniques (e.g., STRIDE~\cite{shostack2014threat}, CORAS~\cite{lund2010model}, attack trees~\cite{saini2008threat}, threat patterns~\cite{abe2013modeling}, PASTA~\cite{wolf2021pasta}, etc.).
Second, favourable performance indicators would result in cost saving for organizations where security experts are scarce.
For instance, STRIDE can be performed using two techniques, per-element and per-interaction.
The documentation of STRIDE-per-interaction includes a larger threat mapping table is considered to be more suitable for expert use and is ``too complex to use without a reference chart handy''~\cite{shostack2014threat}.
Yet, past empirical studies were inconclusive about some performance indicators.
For instance, past controlled experiments have neither found significant difference between STRIDE variants~\cite{tuma2018two} for precision and productivity (whereas a significant difference was measured for recall) nor conducted a statistical analysis of equivalence.


To address the above problems we conducted a replication of a past controlled experiment comparing STRIDE-per-element to STRIDE-per-interaction with 45 student participants.
We have replicated the analysis protocol from~\cite{tuma2018two} and adopted the same measures of success (i.e., precision and productivity).
Similar to the previous study, we found no significant difference (or equivalence) between the treatment groups in precision and productivity.
In addition, we also observe similar variations in the distribution of identified threats per category (more spoofing, tampering, information disclosure and denial of service
compared to repudiation and elevation of privilege threats) for both treatment groups.
In contrast to the original study, we found that per interaction teams performed better (albeit, not significantly) which negates the claim that this technique is more suitable for expert analysts.
Finally, we analyze and discuss the practice of study replication (with human participants) in software engineering. 
The statistical tests used to answer to our hypotheses can be observed in the replication package~\footnote{\url{https://doi.org/10.5281/zenodo.6513366}}.

The rest of this paper is organized as follows. Section~\ref{sec:related-work} discusses the related works and replication best practices. Section~\ref{sec:methods} describes the research questions and methodology of the replication. In Section~\ref{sec:results} we present the results and discuss them in Section~\ref{sec:discussion}.
Section~\ref{sec:threats} lists the threats to the validity of this study. Finally, Section~\ref{sec:conclusion} gives the concluding remarks.

\section{Related Work}
\label{sec:related-work}
We positioned our contributions with respect to existing literature on empirical studies of STRIDE and replication studies in the existing literature on requirements engineering or software design analysis. 
\subsection{Empirical studies of STRIDE}
In addition to the replicated study~\cite{tuma2018two}, several works have investigated STRIDE empirically.

Scandariato et al.~\cite{scandariato2015descriptive} performed a descriptive analysis measuring the productivity, precision, and recall of STRIDE in an academic setting. 
Their study reports similar values for a version of STRIDE-per-element, however the conditions of the descriptive study were different compared to our controlled experiment, therefore the results can not be directly compared.
In comparison to~\cite{scandariato2015descriptive}, this study is replicating a controlled experiment of two STRIDE techniques.

Two studies~\cite{cruzes2018challenges,bernsmed2019threat} conducted case studies investigating the challenges of STRIDE.
Bernsmed et al.~\cite{bernsmed2019threat} conducted semi-structured interviews (with transcription code analysis) with agile organizations and recorded the perceived challenges.
The authors report that practitioners see value in performing STRIDE despite the high manual effort it requires.
Other discovered challenges were related to the lack of expertise by developers conducting the analysis, and uncompatibility of systematic approaches with the Agile workflow.

Stevens et al. \cite{stevens2018CoG} conducted a qualitative case study to investigate the efficacy the Center of Gravity (CoG) technique in an industrial setting.
The CoG, originally conceived as a military strategy, is a risk-first threat analysis technique but has not been extensively used to analyze software security.
The authors designed surveys and classroom sessions and involved 25 practitioners in the study.
Similar to other studies conducted with experts, they report a very high accuracy of the participant results.

\subsection{Replications}
Due to the nature of most software engineering studies, drawing reliable conclusions from a single study might be prone to errors~\cite{shull2004knowledge}. Hence, researchers have resolved to using replication methods as a means of tackling these limitations. According to~\cite{Miller200738} replication would allow the experiment to amass enough “evidence” to be convincing (if  successful). Generally, SE replication studies are directly related to the concept of applying similar experimental procedures as in the original study, on a different participation pool. According to~\cite{santos2019procedure}, this process is aimed at generating new/raw data. However, it is crucial to differentiate between two important concepts, replication and reproduction. In the latter, reproduction, researchers follow replication procedures however, their experimental analysis is performed on the original data with the aim of getting the same results. To this end, reproduction deals with re-analysing raw data obtained from baseline experiments while replications leads to generation of new data that can be combined with other studies to provide joint conclusions~\cite{santos2019procedure}.

Labunets et al~\cite{labunets2013experimental} conducted a controlled experiment that was replicated in~\cite{labunets2017equivalence} using student participants to compare two risk assessment methods, a visual method (CORAS) and a textual method (SREP). The first study found that the visual method was more effective for identifying threats than the textual one. However, the replicated experiment showed that the two techniques were (statistically) equivalent in terms of the quality of identified threats and security controls.
Similar to our work, the authors conduct a TOST analysis to determine statistical equivalence between the treatments.

Jung et al.~\cite{Jung2013223} performed a within-subject experiment to compare two safety analysis techniques, Component Integrated Fault Trees (CFT) and Fault Trees (FT), with 7 phd candidates and replicated the experiment with 11 practitioners. 
Concretely, the authors compare the two techniques in terms of quality of results and perceived method consistence, clarity, and maintainability.
Similar to our work, the authors find no major differences in terms of result quality, and some differences were observed in the perceived measures.
In contrast to our work, Jung et al.~\cite{Jung2013223} compared safety techniques and replicated the experiment with a different population sample (i.e., practitioners).

Several studies have empirically compared~\cite{opdahl2009experimental,karpati2011experimental,karpati2012comparing} and conducted replications~\cite{Jung2013223,Riaz20172127,Rueda2020} requirement engineering techniques (e.g., requirements elicitation). 
For brevity, we direct the interested reader a comprehensive review by Ambreen et al.~\cite{ambreen2018empirical}.

\textbf{Replication best practice.} In order to get a more complete overview of the current status of empirical research in software engineering, we conducted a literature search on some of the major (empirical) software-engineering venues. Our sample consisted of papers form ESEM, IST, TSE and other relevant venues (e.g., RE, JSS, ESEJ, EASE, WICSA). We extracted the literature based on a keyword search on Scopus which leaned 159 results in the domain of computer science. We limited the search to papers investigating requirements engineering or software architecture (design) approaches. We manually examined each paper, with regards to the use of empirical research methods, recruitment of participants (from academia or industry), subject sample size and background, experimental processes (data collection, analysis and processing), statistical tests and results (conclusion). 

\begin{table*}
\caption{Analysis of Replicated Studies. Keyword `same' refers to the replication maintaining the exact measures as the original study, `similar' to slight variations in the measures, `Different' to major variations in the measures, `extra steps' to additional measures and processes being used, and `complements' to support the results and introduce relevant additional information. }
    \centering
    \resizebox{0.9\textwidth}{!}{
    \begin{tabular}{|l|p{1.5cm}|l|l|l|p{2cm}|l|l|l|l|l|p{2cm}|}
    \hline
    Paper & Research Methods (EX, QEX) & RQs & Subjects & Population size& Population Background & Data Collection & Data Cleaning & Data Processing & Statistical Test & Conclusion & Provides Replication Package\\
    \hline
    \hline
    ~\cite{Wang2018} & EX & Similar & Academia & Similar & Same& Same&Same&Same&Additional&Partly Supports&Yes\\
    ~\cite{Abrahao2017144} & EX & Same & Academia & Similar &Different&Similar&Same&Same&Same& Partly Supports& Yes\\
    ~\cite{Fagerholm2019} & EX & Different & Both & Different& Same& Same& Same& Same& Different& Supports& Yes\\
    ~\cite{Riaz20172127} & QEX & Different & Academia & Different &Similar&Similar & Extra Steps &Extra Steps& Same& Partly Supports & No\\
    ~\cite{gomez2014replicated} & EX & Same & Academia & Similar &same& Similar& -& Same& Additional & Partly Supports& No\\
    ~\cite{kosar2012program} & EX & Same & Academia & Different &Same& Same& Same &Similar& Different & Supports &Upon Request\\
    ~\cite{Wagner2019} & EX & Similar & Academia & Different&Same& Same& Same& Same& Different & Complements& No\\
    ~\cite{Jung2013223} & EX & Same & Industry & Similar& Different&Same& Same& Same&Same& Supports& No\\
    ~\cite{Perez-Gonzalez201965} & EX & Similar &Industry & Different& Same &Same& Same& Same&- &Partly Supports& No\\
    ~\cite{Fernandez-Saez2012134} & EX & Same & Academia & Different& Same& Same&Same & Same& Additional & Non conclusive& Yes\\
    ~\cite{Ferreira2022} & EX & Same & Both & Different& Same& Same& Same& Same& Same& Supports& No\\
    ~\cite{Scanniello2014494} & EX & Same & Academia & Same& Same&Extra Steps& Same& Same& Different& Supports& Yes\\
    ~\cite{Scanniello2014} & EX & Same & Academia & Different & Same & Extra Steps& Same& Same&Same & Non conclusive& No\\
    ~\cite{Javed2014215} & EX & Same & Academia & Different& Similar&Same & Same& Same& Same& Supports& No\\
    \textbf{This work} & EX & Similar & Academia & Different & Similar & Similar & Same & Extra Steps & Different & Partly contradicts & Yes\\
    \hline
    \end{tabular}}
    \label{tab:Replicated Studies}
\end{table*}

The resulting analysis was compared to the relevant baseline study and populated in Table~\ref{tab:Replicated Studies}. In total, we analysed 15 studies (which included replication and family of experiments) and limited our analysis to papers produced in the last 10 years. From the analysis we noticed a few trends. 

We could observe two type of replications: close replications and differentiated replications~\cite{lindsay1993design}.
Close replications recreate the same conditions of the original study with the aim of determining to what extent (if at all) the original study is replicable.
On the other hand, differentiated replications include deliberate variations of the original conditions or major effects, which allows researchers to observe the impact of treatment variation on the study outcomes.
First, close replications ~\cite{Wang2018} ~\cite{Abrahao2017144} ~\cite{Riaz20172127} ~\cite{gomez2014replicated} ~\cite{Perez-Gonzalez201965} can only partially support/contradict the original study. 
Indeed, the issue with close replication studies in software engineering was already observed by Lung et al.~\cite{Lung2008191}. 
Instead, replications that go one step further ~\cite{Fagerholm2019} ~\cite{kosar2012program} ~\cite{Wagner2019} ~\cite{Jung2013223} ~\cite{Ferreira2022} ~\cite{Scanniello2014494} ~\cite{Javed2014215}, by either changing the way they analyse (using statistical tests) or analyze from different perspective tend to either support or contradict completely. 

Second, we noticed that, on average, less studies provide replication packages. Availability of replication packages for experiments encourages better replications and complementary studies~\cite{shull2004knowledge}. However, availability of information packages and transfer are still an open question in SE ~\cite{solari2018content}. Particularly for studies involving human participants, researchers are often additionally constrained to share the complete data set due to privacy concerns.

Third, most studies (73\% of our analysed papers) use only academic participants. Although ~\cite{labunets2013experimental},~\cite{svahnberg2008using} agree that students are well suited to perform empirical studies,~\cite{Falcao2015128} argues that the low proportion of professionals used in SE experiments may reduce experiments realism by inhibiting the understanding of industrial software processes. Fourth, many studies do not report important information (such as participant reward, target population, hypothesis, and conclusion and construct validity), as also observed in~\cite{Falcao2021330}.

While ~\cite{jedlitschka2008reporting} outlines the guidelines for reporting controlled experiments studies, the results shown in this research is a clear indication that some aspects of reporting need improvement. However, we understand that some limitations (space, number of pages) are often a determining factor to the amount of information included in reports. None-the-less, it is important to include necessary information to facilitate better understanding and future replication efforts (either internal or external).

\section{Research Methods}
\label{sec:methods}
This work is a replication study of the controlled experiment in~\cite{tuma2018two}. This replication follows the experimental procedure of the original study. The primary goal of the present study is to examine whether the hypotheses in the previous study are upheld given a different participation pool. As a secondary goal we also aim to observe how experiments with threat analysis techniques using human participants are replicated.
\subsection{Research questions}
Our main motivation for conducting this study is to provide a reference direction for researchers interested in undertaking controlled experiments in the field of threat modeling using STRIDE by measuring two dependent variables, productivity and precision. According to \cite{abramo2014you} productivity is the quintessential indicator of efficiency in any production system. Similarly, in this study, productivity is defined as the measure of correctly identified output (True positive) produced within a specified time-frame. In practice, the STRIDE threat modeling process is time-consuming \cite{scandariato2015descriptive}. Any organization or team that is new to threat modeling may be overwhelmed with information on STRIDE application and potentially miss out on important aspects to consider. Even experienced threat modelers may find this method more time consuming and thus adversely affecting productivity. However, \cite{krishnan2017hybrid} claims that providing a structured approach to threat modeling, similar to the one presented in this paper, can be used to mitigate time constraints. On the other hand, measuring precision, defined as correctness of analysis, is crucial in maintaining high productivity. To this end, there is a direct correlation between high productivity and high precision and vice versa. For instance, It would be impractical for a team to have high precision if they have a high number of false positives (\textbf{TP< FP}). 
From a scientific perspective, both productivity and precision obtained in the previous study will be compared to this replication in an attempt at understanding how a different participation pool perceives and handles similar tasks. The difference, or lack thereof, of statistical significance in productivity and precision between the two studies will be used inform the practicability of using non-experts in practical implementation of STRIDE. 
To achieve the aforementioned objectives, this study addresses the following research questions (RQs):

RQ1:\textit{Productivity.} Are the means of productivity between treatment groups equivalent? 

The following null and alternative hypotheses were formulated:

$Hprod_0$: \textit{The productivity difference between the two group means is outside the equivalence interval }

$Hprod_1$: \textit{The productivity difference between the two group means is within the equivalence interval} 

RQ2:\textit{Precision.} Are the means of precision between treatment groups equivalent?

The following null and alternative hypotheses were formulated:

$Hprec_0$: \textit{The precision difference between the two group means is outside the equivalence interval}

$Hprec_1$: \textit{The precision difference between the two group means is within the equivalence interval}


\subsection{Design of the replication}
The following sub-sections will describe how the replication study was conducted. We followed best practices for replicating experiments with human participants (discussed in Section~\ref{sec:related-work}) and tried to stay as close as possible to the original design. For instance, we have adapted a similar training and used the resources published in the original study companion web-site\footnote{\url{https://sites.google.com/site/empiricalstudythreatanalysis/home?authuser=0}}.

\subsubsection{Compared techniques} Threat modeling is a systematic approach performed by security analysts (and domain experts) to identify and mitigate security threats early-on in the design phase of the Software Development Life-cycle (SDL). It includes eliciting the threats, finding and discussing the security attacks, and prioritizing the identified threats to plan for their mitigations.
In this work, we investigate Microsoft's threat modeling methodology, STRIDE~\cite{shostack2014threat}.
STRIDE is an acronym representing six different threat categories which are used to help the analysts in findings the security threats. 
The threat categories, spoofing, tampering, repudiation, information disclosure, denial of service and elevation of privilege have been described further in the documentation of STRIDE.

\begin{figure*}
    \caption{High Level DFD (Home monitoring System)}
    \centering
    \includegraphics[width=0.7\linewidth]{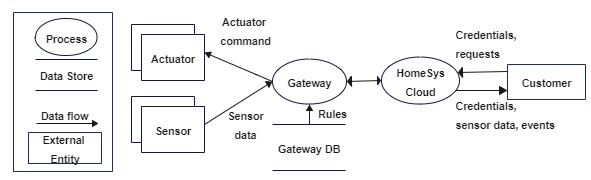}
    \label{fig:High-level DFD}
\end{figure*}
STRIDE is a model-based approach and uses the Data-Flow Diagram (DFD) to represent the software system under analysis. 
A DFD is a graphical model representing the sources of data, the data flows through the software system, and where it is consumed (e.g., at rest). 
Thus, the DFD notation consists of data flows, external entities, processes, data stores, and trust boundaries. The latter are used to separate DFD elements depending on the level of privilege.
For instance, Figure~\ref{fig:High-level DFD} is a DFD representation of how information moves around in a software-based system (in this case, the home monitoring system). 
Analysts can perform the analysis with STRIDE using two variants, per-element and per-interaction \cite{shostack2014threat}. The main difference between the variants is the strategy of visiting the diagram (see figure \ref{fig:High-level DFD}). 
For more details see Appendix~\ref{appendix:a}.

\subsubsection{Description of the Case}
The home monitoring system (HomeSys) is an automated surveillance system designed for residential places. Its main objective is to enable the home-owner to remotely monitor their property. One of the key features of HomeSys is its ability to send notifications when a critical event occurs (eg. sudden spike in temperature, or smoke). The architectural documentation of a HomeSys consists of three component diagrams:
\begin{itemize}
\item Context diagram
\item Gateway component diagram
\item Cloud component diagram
\end{itemize}

The context diagram consists of three external entities, the mote, sensor, and actuator (as also depicted on the DFD in Figure~\ref{fig:High-level DFD}), that communicates with the system. The customer, can access the HomeSys cloud services (depicted as process `HomeSys Cloud' in Figure~\ref{fig:High-level DFD}) via a browser or a mobile application. Using their user profiles, customers can manage their HomeSys installation and gain access to historical data on the dashboard. The Gateway component (depicted as process `Gateway' in Figure~\ref{fig:High-level DFD}) is a hardware device that enables automation of smart homes. 

To properly understand the Home monitoring system, a detailed documentation was made available to the participants. 
\subsubsection{Participants}
All participants are Computer Science Master students enrolled in a course taught by the experimenters. At the beginning of the course an entry survey was conducted to help the experimenters understand participants' background and areas of expertise relevant to the study. Majority of the students had a background in computer science with specific areas of expertise including artificial intelligence, internet and web technologies, software engineering, and computer systems. The remainder (20\%) of the participants had a cyber-security background. Most of them stated that they were new to secure design techniques (e.g STRIDE, threat modeling, Data Flow Diagrams, misuse cases, attack trees etc). Lastly, a large number of the participants rated their familiarity with specific software design models (e.g sequence, component and deployment diagrams) as being very limited. However, the course material also included training lectures on relevant technical skills like STRIDE as a threat modeling tool and components of a DFD.

\subsubsection{Treatment}
First, the students were randomly divided into two groups (A and B) representing the two treatment groups. Group A was tasked with analysing the HomeSys app using STRIDE Per Element while Group B was tasked to use STRIDE per Interaction. Next, participants within each treatment group were randomly assigned a partner. 

\subsubsection{Task}
The experiment started after the participants watched all the necessary training videos assigned to each treatment group. The participants were then asked to complete three tasks: 

\begin{enumerate}
    \item Create a detailed DFD (Figure \ref{fig:High-level DFD} is an example of a high level DFD of the HomeSys application) 
    \item Identify all the possible security threats
    \item Map the threats to their element type
\end{enumerate}

\begin{table}
\caption{CSV Threat Template}
\resizebox{0.5\textwidth}{!}{
\begin{tabular}{|l|l|l|l|l|l|}
\hline
\textbf{ThreadID} & \textbf{ElementName} & \textbf{ElementType} & \textbf{Category} & \textbf{Description} & \textbf{Assumption} \\ \hline
1                 & External Entity X    & External Entity      & Spoofing          & ......               & .....               \\ \hline
2                 & Process X            & Process              & Repudiation       & .....                & .....               \\ \hline
\end{tabular}
\label{tab:threat-template}
}
\end{table}
In the first task, participants were instructed to refer to the architectural description of the HomeSys monitoring system and use UML diagrams to create DFDs (similar to Figure~\ref{fig:High-level DFD} ). When participants were satisfied with their DFDs, they began analysing it using the different STRIDE variant depending on the treatment group. Using STRIDE per-element, participants analysed every element independently (e.g., starting in the upper-left corner of Figure~\ref{fig:High-level DFD} ). To facilitate this analysis, mapping element types to threat categories, STRIDE also provides a mapping chart (see \figureautorefname{ 1}). On the other hand, STRIDE per-interaction group would visit every interaction in their DFD to elicit a possible threat scenario. An example of an interaction using Figure~\ref{fig:High-level DFD}  is when an actuator sends a command to the gateway. For the third task, the participants received a threat template (see Table~\ref{tab:threat-template}) where they categorised each threat based on its element name, type, threat category and a description of how the threat might occur. Optionally, the students could state their assumptions. After completion, the participants were required to upload a copy of their DFD accompanied by a filled-in template with the identified security threats.

Time taken to complete the task was captured using an online survey tool. The start time was entered when the participants started the survey and ended after the submission of the exit questionnaire.
\subsubsection{Measures}
To analyse the participants submissions, a ground truth was developed. The process of creating a ground truth was divided into two sub-tasks. 
In the first one, one experimenter analysed the ground truth from the previous experiment (each unique DFD had a different ground truth in the previous experiment) to consolidate a single list of threats that are applicable in the HomeSys monitoring system. Secondly, the ground truth was discussed between the authors to come up with a final draft. The final ground truth was used as a guide to mark correct and incorrect threats. Correctly marked threats were categorised as True positives while incorrect threats were considered false positives. Unlike the previous experiment, we do not identify False negatives and therefore recall was not analysed. Table~\ref{tab:threat classification} is a visual representation of how threat classification was done. In the first row, the participants reported threats were compared to the ground truth. The other rows checked the consistency within the participants reported threats.

\begin{table}
\footnotesize
\caption{Criteria for Threat Classification}
\begin{tabular}{|p{0.45\columnwidth}|p{0.45\columnwidth}|}
\hline
\textbf{True Positive} & \textbf{False Positive} \\ \hline
Correct element name, type, and STRIDE category  & Correct element name, type, but wrong STRIDE category \\ \hline
Description matching threat category       & Mismatch of description and threat category      \\ \hline
Assumptions matching the description & Mismatch of  assumption and description   \\ \hline
Clearly stated description and assumptions & Lack of either description or assumption or both \\ \hline
\end{tabular}
\label{tab:threat classification}
\end{table}
Two measures of success were used to determine the outcome of the experiment, team productivity and precision.\\
Productivity- in measuring the productivity of each STRIDE variant, the sum of each teams' TP were was divided by the time taken (in hours) to complete tasks (see tasks). The formulae for productivity is defined below. \\

  $$ Prod=\frac{TP}{Hour}$$

Precision- to determine the precision, the rates of True Positives and False Positives were considered. Using both metrics, precision was calculated using the formulae below.

$$Prec= \frac{TP}{TP+FP}$$

\subsubsection{Execution}
\begin{table}[]
 \caption{List of all Available Materials}
 \resizebox{0.5\textwidth}{!}{
\begin{tabular}{|c|c|c|}
\hline
\textbf{Element} & \textbf{Interaction} & \textbf{Both Groups} \\ \hline
\begin{tabular}[c]{@{}c@{}}Training video: \\ STRIDE-per-element\end{tabular} & \begin{tabular}[c]{@{}c@{}}Training video:\\ STRIDE-per-interaction\end{tabular} & \begin{tabular}[c]{@{}c@{}}Training Video: \\ HomeSys monitoring system\end{tabular} \\ \hline
\begin{tabular}[c]{@{}c@{}}STRIDE book: \\ Chapter three (element)\end{tabular} & \begin{tabular}[c]{@{}c@{}}STRIDE book: \\ Chapter three (interaction)\end{tabular} & \begin{tabular}[c]{@{}c@{}}HomeSys \\ Documentation\end{tabular} \\ \hline
\end{tabular}
\label{tab:Available materials}
}
\end{table}

Training- Prior to the start of the experiment,the participants were given access to comprehensive training videos the day before via a learning management system and instructed to watch them. The training materials included a description of the HomeSys monitoring system, a STRIDE training video which was different depending on the group (group A received STRIDE per Element). Additional documentation was also provided, Table~\ref{tab:Available materials} is a detailed list of all materials provided to each treatment group.

Physical Labs- The experiment was conducted during a two hour class where the different teams were allocated different labs depending on their treatment group. Due to the large number of teams (22 pairs), the experimenters separated the groups into four labs with each lab being supervised by either a teaching assistant or the course lecturers. The participants were allowed to ask the supervisors questions but team interactions were limited to partners.

Reports- All necessary metrics, such as duration taken to complete the task, the group member who submitted the report, and the uploaded DFDs and CSV templates were captured on Qualtrics. The last step in the experiment consisted of an exit questionnaire which consisted of questions aimed at gauging the participants experience, familiarity, confidence on their performance, and general knowledge on the subject. The responses were captured on a Likert scale.



\section{Results}
\label{sec:results}
In addition to the analysis performed in~\cite{tuma2018two}, if we can not find significant difference, we also check whether the location shift between the two sample means in precision is statistically equivalent.
For both measures (productivity and precision) we performed a T-test equivalence test to check whether the location shift between the two sample means in the treatment groups is statistically equivalent. The delta was estimated by the authors due to lack of larger samples from other similar studies that could be used to empirically derive the delta estimate. For productivity, we consider a delta of 0.5 correct threats per hour as equivalent, while for precision we consider a delta of 5\% as equivalent.

The outcome of the study will be presented in this section by answering the research questions.
\subsection{True Positives and False Positives}

\begin{table}[]
\caption{Summary Observations in Per-Element Group}
\resizebox{0.4\textwidth}{!}{
    \centering
    \begin{tabular}{|c|c|c|c|c|c|}
    \hline
    Element Group (ID) & Time Taken (hours)& TP & FP & Productivity & Precision\\
    \hline
    \hline
    0 & 1.20& 3 & 10 & 2.49 & 0.23\\
    1 & 1.65& 4 & 6 & 2.41 & 0.40\\
    2 & 1.67& 2 & 6 & 1.19 & 0.25\\
    3 & 1.56& 7 & 13 & 4.47 & 0.35\\
    4 & 1.70& 7 & 1 & 4.09 & 0.87\\
    5 & 1.90& 7 & 3 & 3.66 & 0.70\\
    6 & 1.72& 0 & 7 & 0.00 & 0.00\\
    7 &2.40& 10 & 5 & 4.15 & 0.66\\
    8 & 2.49& 5 & 5 &2.00 & 0.50\\
    9 & 3.16& 9 & 6 &2.84 & 0.60\\
    10 & 5.48& 3 & 10 &0.54 & 0.23\\
    11 &2.27& 7 & 63 &3.08 & 0.10\\
    Mean & 2.27& 5.33 & 11.25 & 2.5 & 0.40\\
    Standard Deviation & 1.13& 2.99 & 16.61 & 1.19 & 0.26\\
    \hline
    \end{tabular}}
    \label{tab:element summary}
\end{table}

\begin{table}[]
\caption{Summary Observations in Per-interaction Group}
\resizebox{0.4\textwidth}{!}{
    \centering
    \begin{tabular}{|c|c|c|c|c|c|}
    \hline
    Interaction Group (ID) & Time Taken (hours)& TP & FP & Productivity& Precision\\
    \hline
    \hline
    0 & 1.93& 9 & 5 & 4.65 & 0.64\\
    1 & 0.92& 10 & 30 & 10.81 & 0.25\\
    2 & 2.35& 1 & 5 & 0.42 & 0.16\\
    3 & 1.94& 8 & 4 & 4.10 & 0.66\\
    4 & 1.90& 9 & 2 & 4.71 & 0.81\\
    5 & 1.91& 8 & 4 & 4.17 & 0.66\\
    6 & 1.92& 4 & 27 & 2.07 & 0.12\\
    7 & 2.02& 0 & 11& 0.00 & 0.00\\
    8 & 2.08& 9 & 5 & 4.30 & 0.64\\
    9 & 2.07& 1 & 2 & 0.48 & 0.33\\
    Mean & 1.91& 5.90 & 9.50 & 3.57 & 0.43\\
    Standard Deviation & 0.37& 3.95 & 10.34 & 3.17 & 0.28\\
    \hline
    \end{tabular}}
    \label{tab:interaction summary}
\end{table}

\begin{figure*}
    \centering
    \includegraphics[width=0.9\linewidth]{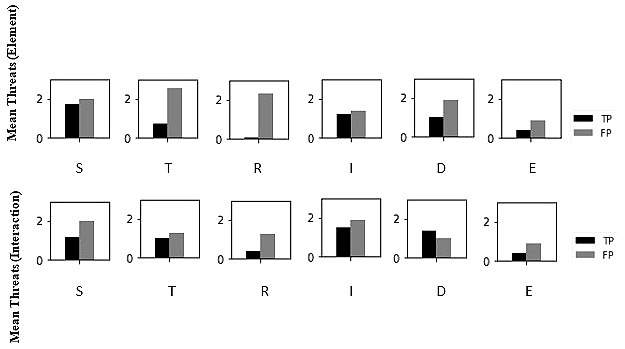}
    \caption{Precision Per Threat Category (per-element)}
    \label{fig:threats per category}
\end{figure*}

Table~\ref{tab:element summary} and \ref{tab:interaction summary} show a summary of all necessary observations in both groups (Identified TP, FP, overall productivity, and precision for each team). The average number of True positives identified by the per-Element group was \textbf{5.33} (\textit{STDEV=} \textbf{2.99}) with a corresponding average False positive of \textbf{11.25} (\textit{STDEV=}\textbf{16.61}). The per-Interaction group reported an average of \textbf{5.90} (\textit{STDEV=} \textbf{3.95}) True Positives and \textbf{9.50} (\textit{STDEV=}\textbf{10.34}) false positives. Figure~\ref{fig:threats per category} is a representation of the average TP and FP of each threat category across the treatment groups. To understand the high occurrence of FP in both treatment groups, we further analysed the submitted threat templates. We discovered that at least one team in each treatment group listed a huge number of threat categories without a corresponding description, assumption, or both despite repeated warning during the training to avoid making this mistake. According to the criteria for threat classification presented on Table~\ref{tab:threat classification}, such mistakes automatically led to the classification of some threat categories as false positives regardless of whether or not they were correctly matched to the element name and element type. The authors concluded that the members of these two teams lacked an understanding of what the task required them to do. While the teams performed relatively better in identifying spoofing, information disclosure, and denial of service threats, the high frequency of FPs were attributed to the mistake mentioned above.  However, a significant amount of tampering threats identified had been matched with the wrong element type, a common occurrence in threat analysis using non-experts \cite{Shostack08}. Generally, repudiation and elevation of privilege attacks were less identified with an average of two attacks per team. For teams that listed more than two repudiation or elevation of privilege attacks, they either lacked a threat description, an assumption or both.

\subsection{RQ1: Productivity}
The study defined productivity as the number of correctly identified threats (True Positives) per time taken to complete the task (in hours). True positives were also used to represent the amount of work (effort) performed by a team. The average time spent by the teams performing STRIDE per element is 2.27 h (\textit{STDEV=1.13}), while STRIDE per interaction teams spent an average of 1.9 h (\textit{STDEV= 0.37}) (not statistically significant).The overall productivity (calculated as \textbf{TP/Hrs}) was dependant on the amount of correctly identified threats. The higher the TP the higher the average productivity, and vice versa. As depicted in \tableautorefname{ 5}, the average productivity in the per-element group is \textbf{2.5 \textit{TP/h}} (\textit{STDEV=} \textbf{1.19}), and per-interaction teams is \textbf{3.57 \textit{TP/h}} (\textit{STDEV=}\textbf{3.17}). The equivalence test on productivity produced a 0.7159 p-value, supporting the null hypothesis ($Hprod_0$). Therefore, we conclude that the groups productivity is not equivalent.

\subsection{RQ2: Precision}
\begin{figure}
    \centering
    \includegraphics[width=0.9\linewidth]{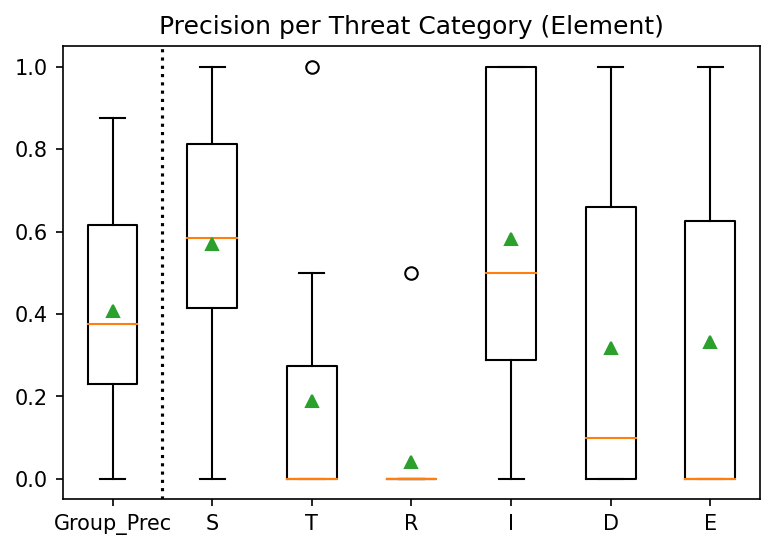}
    \caption{Precision Per Threat Category (per-element)}
    \label{fig:boxplot element}
\end{figure}

\begin{figure}
    \centering
    \includegraphics[width=0.9\linewidth]{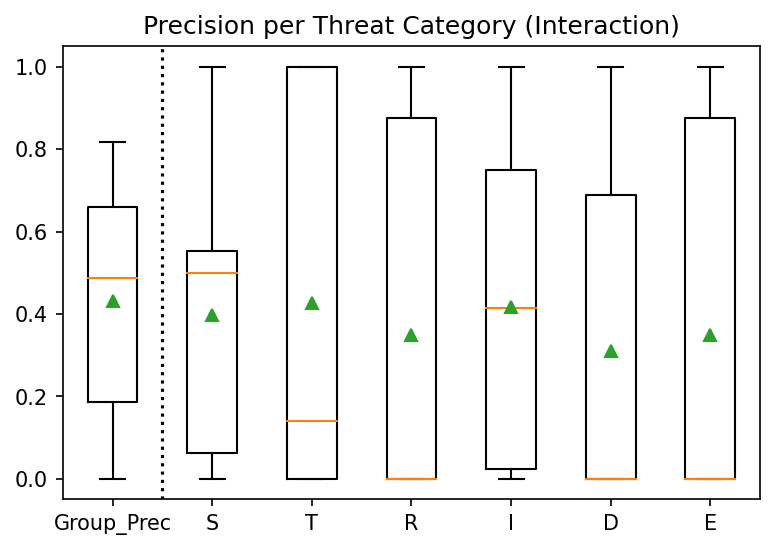}
    \caption{Precision Per Threat Category (per-interaction)}
    \label{fig:boxplot interaction}
\end{figure}

The study defined precision as the correctness of the analysis calculated as \textbf{\textit{TP/TP+FP}}. Overall, the average precision for each treatment group was \textbf{0.43} (\textit{STDEV=} \textbf{0.28}) for per-interaction and \textbf{0.40} (\textit{STDEV=}\textbf{0.26}) for per-element. Similar to the original study, there is no statistically significant difference between the precision of both teams. However, we conducted an equivalence test (\textit{p-value=0.3986}), since the p-value is outside the equivalence interval, supporting the null hypothesis ($Hprec_0$), we concluded that the location shift between the two sample means of precision from both treatment groups is statistically not equivalent.  Figure~\ref{fig:boxplot element} and \ref{fig:boxplot interaction} display precision for each threat category (median-yellow line, mean-green triangle).


\subsection{Exit Questionnaire}
\label{subsec:Exit}
The final step in the experiment was conducting an exit questionnaire, meant to gauge the students perceived complexity in performing the task. Their responses were recorded using a predefined likert scale. The training materials made available to the participants was instrumental in improving their familiarity with the home monitoring system (HomeSys), evident from the exit questionnaire. Both treatment groups found the process of building a DFD neither challenging nor easy, which also reflected on the quality of the DFD they uploaded. Analyzing the DFD and threat mapping using the different variant charts was considered to be easy by both teams. However, participants strongly agreed that the process was time consuming and tedious. There was a slight difference in how much each treatment group liked their technique. Majority of the teams in Per-element chose "I do not like it" while Per-interaction teams stated that they "quite liked it", which can be used to explain their slightly better productivity.
   

\section{Discussion}
\label{sec:discussion} This section reports analysis from comparing our results to the original study.  

\textbf{Comparison of True Positives and False Positives.} The original study \cite{tuma2018two} reported slightly better averages of TP from Per-element group (14.3) compared to per-interaction (11.6). However, the average FPs recorded were relatively lower for per-element (14.2) while per-interaction had 15 FP. In comparison, our study reported lower averages across both teams (per-element: 5.33 TP and 11.25 FP vs per-interaction: 5.90 TP and 9.50 FP). Overall, the original study produced higher averages across both treatment groups. We conclude this was a result of the original study having relatively higher participants per team (4-5 students) while the replication had 2 students per team. However, analysis done on both studies show that there were no significant differences between the average number of TP and FP across treatment groups.

\textbf{Comparison of productivity.} As discussed in Section~\ref{sec:results}, the average time spent performing the tasks in the replication was not statistically different, similar to the original study from~\cite{tuma2018two}. Replication (per-element: 2.27 h, per-interaction: 1.9 h), original (per-element: 3.5 h, per-interaction: 3.95 h). The resulting average productivity across both studies were also not statistically different. Replication productivity (per-element: 2.5 TP/h, per-interaction: 3.57 TP/h), original productivity (per-element: 4.35 TP/h, per-interaction: 3.27 TP/h). It is clear that per-interaction group in the replication study performed better than in the original. Productivity is driven by the number of TP, compared to the original study, the productivity was much lower in the replication. In contrast to the past study, in the replication, per-interaction was slightly better (not significant) in terms of productivity. A possible explanation for this is the tendency of per-interaction teams to liking the technique more (as described already in \ref{subsec:Exit})

\textbf{Comparison of precision.}
We also compared the precision of teams from~\cite{tuma2018two} and our treatments groups. Similar to productivity, the resulting precision measure by our replication was similar to the precision measured in the past study. Replication precision (per-element: 0.40, per-interaction: 0.43), original precision (per-element: 0.62, per-interaction: 0.58). Although the difference is not significant, per-interaction teams in the replication study were slightly more precise than per-element (again, the reverse is true for the original study). 


\textbf{Comparison of threat categories.} To complete our analysis, we compared the average number of identified threats per category across each treatment group in the original and replication studies. On average, both treatment groups identified more (TPs and FPs) spoofing, tampering, information disclosure and denial of service compared to repudiation and elevation of privilege threats. This trend was also observed in other studies \cite{tuma2018two}\cite{scandariato2015descriptive}. We concluded that such threats are easily identifiable by participants regardless of the treatment group. On the other hand, repudiation and elevation of privilege threats were less identified (averagely lower TP). A possible explanation for this might be that students lack a clear understanding of how vulnerabilities can be exploited to result in repudiation or elevation of privilege.

\section{Threat to validity}
\label{sec:threats}
This section will discuss the various limitations with the experiment that might pose a threat to validity to the results. 

\textbf{\textit{Participant selection and sample size.}} The use of student participants as opposed to experts might be considered a treat to validity due to their limited knowledge of implementing industrial-level techniques. However, \cite{labunets2013experimental} states that student participants are well-suited for empirical experiments. The sample size, 45 students, was relatively smaller than in the original study (110 students). Since the experiment was conducted in-person, the authors were satisfied with this sample size because it was easy to manage given the Covid-19 mandates on social distancing in public gatherings.

\textbf{\textit{Participants' Motivation.}} Participation in the experiment was part of the learning objectives of the course, however only participation was graded (as a pass or fail) and not their performance in the task of the experiment. This was an important decision because alternatively, students tend to submit more security threats (which results in many false positives) to maximize their grade. Second, the students actually used the data they produced as part of an assignment in the course. Hence the students were motivated to undertake the study seriously.

\textbf{\textit{Group Composition and Execution.}} Compared to the previous study whereby teams were made up of 3-5 participants, our replication consisted of two members per team with an exception of one team (3 members). The reduced team composition might present an unequal distribution of knowledge of task or techniques used in the experiment. However, we provided access to all training materials 24 hours prior to the experiment. Participants could access these materials at anytime, even during the experiment. Additionally, there was at-least one lecturer or teaching assistant present in the labs to answer any questions that might arise. 
We remark that the time given to participants to perform the task in future replications should be proportional to the system under analysis (in our replication we allotted about 2h for the analysis, whereas the original study allotted about 4h). 




\section{Conclusion}
\label{sec:conclusion}
This study focuses on threat modelling techniques by investigating performance indicators (productivity and precision) of two STRIDE variants (Element and Interaction). We conducted a replication of the original study ~\cite{tuma2018two} in a controlled experiment. Our participants are enrolled Master's students with a background in computer science and cyber-security. The contribution of this work is as follows:
\begin{enumerate}
    \item A quantitative comparison of performance indicators of STRIDE by testing the hypothesis of the original study on a different participation pool. From the results, study found that per-interaction teams had better productivity and precision compared to per-element while in the original study per-element performed better in both measures.    
    \item A comparative analysis of SE replicated studies with the aim of providing suggestions for future replication designs. For the best practices of replication, we analysed 15 technical papers obtained from some of the major software-engineering venues, and documented the trends identified. We also explained how these trends may be a hindrance to performing complete (and successful) replications.
\end{enumerate}

As future work, we plan to conduct similar studies with industry professionals with (possibly) prior experience in threat analysis. 
In addition it would be interesting to observe performance (and perceived performance) differences between expert to non-expert populations.
Finally, the target population of our study (and related studies investigating STRIDE) was computer science students (or professionals in other studies), however, such a sample is not very diverse e.g., in terms of gender.
We are interested to observe whether diversity factors plays a role in security threat analysis and to what extent biased decision making (if any) is more pronounced in the computer science domain when compared to populations with different educational backgrounds (e.g., with a primary background in social sciences).

\section*{Acknowledgements} 
Witheld for anonymous submission

\bibliographystyle{ACM-Reference-Format}
\bibliography{sample-base}

\appendix
\section*{Appendix A: STRIDE-per-element vs per-interaction}
\label{appendix:a} In addition to DFD creation, analysts often make assumptions about the system under analysis.
Assumptions are statements (which may or may not be true) about the domain, functionality and security of its' components.
They can be identified when the diagram is built and they can arise during the analysis.
Landuyt \cite{van2021descriptive} defines threat assumptions as information used to hypothesise certain system properties that are relevant to the attack identified. Threat assumptions are important in threat analysis as they can be used to justify threat existence and prioritise mitigation efforts.

After the threats are elicited, analysts discuss attack scenarios and document them with threat descriptions.
A threat description contains concrete steps of the identified security threats and how they may impact system components associated to it. 
Finally, the threat can be prioritized according to estimation of risk (this step is not investigated in our work).



\textbf{STRIDE-per-Element.} This technique is performed by analysing each component in a DFD individually. 
To limit effects of threat explosion, the documentation of STRIDE provides a threat to element mapping table.
For each element type a different subset of threat categories are suggested.
For instance, for external entities (such as ``Sensor'' in Figure~\ref{fig:High-level DFD}),~\cite{shostack2014threat} suggests investigating spoofing and repudiation threats.
STRIDE-per-element is described as a simplified approach to identifying threats that can be easily understood by the beginner~\cite{shostack2014threat}.

\textbf{STRIDE-per-Interaction.} However, threats can not be always discussed in a vacuum as they are a result of some interaction in the system. An interaction is a tuple of origin, destination, and the interaction itself (e.g., the ``Customer'' sending ``credentials'' to the ``HomeSys Cloud'' in Figure~\ref{fig:High-level DFD}).
In contrast to STRIDE-per-element, this technique is performed by systematically visiting each interaction in a DFD. For instance, when an external entity, such as ``Sensor'', is passing information to a process, such as ``Gateway'' with no logging in place, the Gateway is able to deny having received sensitive information from the Sensor. In this case, ~\cite{shostack2014threat} suggests investigating repudiation threats. Again, to limit the number of threats that the analysts need to  consider, a threat to interaction mapping table is used~\cite{shostack2014threat}.
STRIDE-per-interaction is considered to be more suitable for expert use and is ``too complex to use without a reference chart handy''~\cite{shostack2014threat}.


\end{document}